\documentstyle[galley]{mn}
\twocolumn
\input{epsf}
\newcommand{\bm}[1]{{\bf#1}}

\def\Real{{\rm I\mathchoice{\kern-0.70mm}{\kern-0.70mm}{\kern-0.65mm}%
  {\kern-0.50mm}R}}

\font \bolditalics = cmmib10
\def\bx#1{\leavevmode\thinspace\hbox{\vrule\vtop{\vbox{\hrule\kern1pt
        \hbox{\vphantom{\tt/}\thinspace{\bf#1}\thinspace}}
      \kern1pt\hrule}\vrule}\thinspace}

\def \vc #1{{\textfont1=\bolditalics \hbox{$\bf#1$}}}

  \def\xg{{\bf x}} 
   
\def\kg{{\bf k}}   
   \def\sg{{\bf s}}
  \def\nablag{{\vc \nabla}}
  
 \def\thetag{{\vc \theta}}   \def\gammag{{\vc \gamma}}

   \def\Ac{{\cal
A}}   
         
\def\be{\begin{equation}} \def\ee{\end{equation}}
\def\ba{\begin{eqnarray}} \def\ea{\end{eqnarray}}

\title{Weak Lensing Predictions at Intermediate Scales}

\author[L. Van Waerbeke et al.]  {Ludovic Van Waerbeke$^{1,2}$, Takashi
Hamana$^2$, Rom\'an Scoccimarro$^3$,\cr Stephane Colombi$^{2,4}$, Francis
Bernardeau$^5$ \\ $^1$ Canadian Institute for Theoretical
Astrophysics, 60 St George Str., Toronto, M5S 3H8, Canada\\ $^2$
Institut d'Astrophysique de Paris, 98$^{\rm bis}$ Bld Arago, 75014
Paris, France\\ $^3$ Institute for Advanced Study, Einstein Drive,
Princeton, NJ 08540, United States\\  $^4$ Numerical Investigations
in Cosmology (N.I.C.), CNRS, France\\ $^5$ Service de Physique
Th\'eorique. C.E. de Saclay. 91191 Gif sur  Yvette Cedex, France\\ }

\pagerange{000,000}

\begin{document}

\maketitle
\begin{abstract} 
As pointed out in previous studies, the measurement of the skewness of
the convergence field $\kappa$ will be useful in breaking the
degeneracy among the cosmological parameters constrained from weak
lensing observations. The combination of shot noise and finite survey
volume implies that such a measurement is likely to be done in a range
of intermediate scales ($0.5'$ to $20'$) where neither perturbation
theory nor the hierarchical ansatz apply. Here we explore the behavior
of the skewness of $\kappa$ at these intermediate scales, based on
results for the non-linear evolution of the mass bispectrum. We
combined different ray-tracing simulations to test our predictions,
and we find that our calculations describe accurately the transition
from the weakly non-linear to the strongly non-linear regime. We show
that the single lens-plane approximation remains accurate even in the
non-linear regime, and we explicitly calculate the corrections to this
approximation.  We also discuss the prospects of measuring the
skewness in upcoming weak lensing surveys.
\end{abstract}


\section{Introduction}

Weak lensing by large scale structures has been recognized for a
decade as a potentially powerful tool to probe the distribution of
dark matter in the Universe, as well as a way to measure the
cosmological parameters
(\cite{B91,ME91,K92,V96,BWM97,JS97,S98}). Reports of sound detections of
such an effect have been done recently by different groups
(\cite{VW00,WT00,BRE00,KWL00}) from independent data sets and data
analysis procedures.  These first measurements are detections of the
excess variance of galaxy ellipticities with respect to a random
orientation of the intrinsic ellipticities, and are in very good
agreement with theoretical expectations based on cluster normalized
models of structure formation. These detections demonstrate that the
systematic errors can be controlled down to a low enough level; this
is a remarkable achievement that paves the way to genuine
reconstruction of projected mass maps of the Universe.

The aim of projected mass maps is not only to measure the projected mass
power spectrum. It has indeed been stressed that valuable constraints
on cosmological parameters are expected to come from a joint
measurement of variance and higher order moments such as the skewness
of the convergence (\cite{BWM97,JS97,S98,WBM99}). This is because the
variance is strongly dependent on both the amplitude of the mass power
spectrum and the density of the Universe, while the skewness (properly
normalized) is essentially a measure of the latter.  Another
motivation for measuring the skewness of the convergence comes from
recent works on the possible existence of intrinsic correlations in
the galaxy ellipticities (\cite{HRH00,CM00,CKB00}). This might indeed
challenge the cosmological interpretation of weak lensing surveys if
part of the signal is due to intrinsic alignment of galaxies. However,
additional information on the statistics of the signal such as the
skewness can provide a way to confirm the nature of it (e.g. a signal
dominated by intrinsic ellipticity alignments may not lead to a
measurable skewness as pointed out in \cite{CM00}).

Current skewness predictions rely mostly on perturbative analysis of
structure growth which is expected to break down at scales (around a
few arcmins) where the lensing signal is easier to measure. It is
therefore desirable to dispose also of intermediate and small scale
predictions for the higher order moments. For the second moment, the
variance, the nonlinear evolution of the power spectrum have been used
since the early calculations (\cite{ME91,JS97,S98}). These calculations
are based on a fitting formula for the non-linear evolution of the
mass power spectrum (\cite{H91,PD96}) accurate to $\sim 15\%$, depending
on the cosmological model.  \cite{JS97} and \cite{S98} used this
phenomenological approach to compute the non-linear convergence power
spectrum, the small-scale shear variance and its correlation
functions. The results are in remarkably good agreement with
ray-tracing simulations (\cite{JSW00}).  Unfortunately the non-linear
evolution of higher order statistics cannot be simply described by the
non-linear power spectrum alone because a description of the
non-linear evolution of mode coupling is also required. For instance,
the non-linear evolution of the skewness directly depends on how the
bispectrum evolves.

Prescriptions for the non-linear evolution of the bispectrum were not
available until very recently.  Such an issue is obviously closely
related to the understanding of the non-linear gravitational dynamics
that has been extensively studied for the last 10 years. Although
these efforts were primarily done for studying the galaxy clustering
properties one can take advantage of the results that have emerged in
this domain. Recently, Hui (1999) extended the skewness calculations
in the strongly non-linear regime using ``hyperextended perturbation
theory'' (hereafter HEPT, \cite{SF99}). The predictions agree well
with ray tracing simulations (\cite{JSW00}) at very small scales
($0.1'$), but overestimate the simulation values at larger
scales. This is expected since HEPT was proposed to work in the
non-linear regime, where the hierarchical clustering ansatz is assumed
to hold.  Extensions of this approach presented in the literature
(\cite{V00,MC00,MJa00,MJb00} and \cite{BV00}) are also based on the
hierarchical ansatz, and thus suffer from the same limitations.
Recently, Scoccimarro \& Couchman (2000) obtained a fitting formula
for the bispectrum based on numerical simulations of CDM models. In
this paper, we use this result to compute the skewness of the
convergence $\kappa$ and compare it to ray-tracing simulations. This
allows to bridge the gap between previous results restricted to either
the non-linear regime or the weakly non-linear regime where
perturbation theory holds. A different approach to this problem, using
Press-Schechter halos, is given by Cooray \& Hu (2000).

Taking advantage of these investigations, we also address the
importance of the lens-lens coupling effects and of the Born
approximation, which are routinely used in weak lensing calculations.
It is usually assumed (see \cite{BWM97}) in semi-analytical
predictions that the coupling between lens planes is negligible, as
well as the successive deflections of the lensed light ray. This is
equivalent to assume that the Universe can be approximated by a single
lens plane at some {\it fiducial} redshift.  There is {\it a priori}
no reason to believe that these approximations are still valid in the
non-linear regime. Here we test quantitatively these approximations
down to very small scales both using our ray-tracing simulations and
semianalytic predictions.

The questions we address in this paper are therefore the following:
does the non-linear prediction of the skewness $S_3(\kappa)$ based on
the non-linear evolution of the bispectrum describe correctly
intermediate scales? Is the skewness of the convergence still an
efficient way of discriminating between different cosmologies? Are the
corrections to the unperturbed ray-path given by the Born approximation
and lens-lens coupling terms still negligible at non-linear scales?

The paper is organized as follows.  Basic definitions, perturbative
expressions and the description of the non-linear bispectrum are
reviewed in Section 2. In Section 3, we give the relevant equations to
calculate the variance and the skewness of the convergence, including
correction terms due to lens-lens couplings and deviations from the
Born approximation. Section 4 describes the ray-tracing simulations. 
We present our results in Section 5 and conclusions in Section 6.

\section{Density field statistics}

\subsection{Perturbative Regime}

We assume the standard model of structure growth from initial Gaussian
fluctuations. Perturbation theory (\cite{P80,Fry84}) which describes the
evolution of the density contrast $\delta(\xg,t)=
(\rho(\xg,t)-\bar\rho(t))/\bar\rho(t)$ is well known, therefore we
shall quote here only the results and notations needed in this
paper. The Fourier transform $\tilde\delta(\kg,t)$ of the density
contrast is defined as

\begin{equation}
\tilde\delta(\kg,t)=\int {{\rm d^3}\xg\over (2\pi)^3} \delta(\xg,t)~
e^{-i\kg\cdot\xg}.
\end{equation}
If $\delta \ll 1$ (perturbative regime), for an Einstein-de Sitter
Universe ($\Omega_0=1, \Lambda=0$), the solution of the equation of
motion of $\delta$ can be written as the following perturbation
expansion

\begin{equation}
\tilde\delta(\kg,t)={\displaystyle \sum_{i=1}^\infty}
a^i(t)\tilde\delta^{(i)}(\kg),
\end{equation}
where $a(t)$ is the scale factor such that $a(t_0)=1$ today. For
$\Omega\ne 1$ the time dependence can still be factorized to a very
good accuracy (see eg. \cite{BJCP92} for full second-order results)
although the growth factor is a more complicated function of time than
$a(t)$. At any order, the $\tilde\delta^{(i)}(\kg)$'s can be expressed
in terms of the first order fluctuation $\tilde\delta^{(1)}(\kg)$
\cite{G86}

\begin{eqnarray}
\tilde\delta^{(i)}(\kg)&=&\int{\rm d^3\kg_1}...\int{\rm
d^3\kg_i}~\delta_D(\kg-\kg_1-...-\kg_i)\times\nonumber\\
&&F_i(\kg_1,...,\kg_i)
\tilde\delta^{(1)}(\kg_1)...\tilde\delta^{(1)}(\kg_i),
\label{deltadef}
\end{eqnarray}
where the kernels $F_i(\kg_1,...,\kg_i)$ are non-linear scalar
functions of the wave vectors $\kg_1,...,\kg_i$. The power spectrum
$P(k)$ and the bispectrum $B(\kg_1,\kg_2,\kg_3)$ are defined as

\begin{eqnarray}
\langle\tilde\delta(\kg_1)\tilde\delta(\kg_2)\rangle&=&
\delta_D(\kg_1+\kg_2)~P(k) \cr
\langle\tilde\delta(\kg_1)\tilde\delta(\kg_2)\tilde\delta(\kg_3)\rangle&=&
\delta_D(\kg_1+\kg_2+\kg_3)~B(\kg_1,\kg_2,\kg_3).
\label{spectradef}
\end{eqnarray}
Incorporating Eq.(\ref{deltadef}) into Eq.(\ref{spectradef}) leads to
the expression of the bispectrum as a function of the second-order
kernel $F_2(\kg_1,\kg_2)$ and the power spectrum

\begin{eqnarray}
B(\kg_1,\kg_2,\kg_3)&=&2~F_2(\kg_1,\kg_2)~P(k_1)P(k_2)\nonumber\\
&+&2~F_2(\kg_2,\kg_3)~P(k_2)P(k_3)\nonumber\\
&+&2~F_2(\kg_1,\kg_3)~P(k_1)P(k_3),
\label{bispectredef}
\end{eqnarray}
where (\cite{Fry84})

\begin{equation}
F_2(\kg_1,\kg_2)={5\over 7}+{1\over 2}{\kg_1\cdot\kg_2\over k_1
k_2}\left({k_1\over k_2}+ {k_2\over k_1}\right)+{2\over
7}\left({\kg_1\cdot\kg_2\over k_1 k_2}\right)^2.
\label{F2kernel}
\end{equation}
Note that Eq.(\ref{spectradef}) implies that the bispectrum is defined
only for closed triangles formed by the wave vectors
$\kg_1,\kg_2,\kg_3$.

\subsection{Non-Linear Power Spectrum and Bispectrum}

The non-linear prediction of the skewness requires two ingredients:
the non-linear evolution of the power spectrum and bispectrum. To
describe the non-linear power spectrum, we use the fitting formula of
(\cite{PD96}), based on early work by (\cite{H91}), and obtained by
fitting the non-linear power spectrum in numerical simulations.

In order to describe the bispectrum at all scales, we use the fitting
formula derived by Scoccimarro \& Couchman (2000) for the
non-linear evolution of the
bispectrum in numerical simulations of CDM models, extending previous
work for scale-free initial conditions (\cite{SF99}). The kernel
$F_2(\kg_1,\kg_2)$ in Eq.(\ref{bispectredef}) is simply replaced
by an {\it effective} kernel $F_2^{\rm eff}(\kg_1,\kg_2)$ such that

\begin{eqnarray}
F^{\rm eff}_2(\kg_1,\kg_2)&=&{5\over 7}~a(n,k_1)a(n,k_2)\nonumber\\
&+&{1\over 2}{\kg_1\cdot\kg_2\over k_1 k_2}\left({k_1\over k_2}
{k_2\over k_1}\right)~b(n,k_1)b(n,k_2)\nonumber\\ &+&{2\over
7}\left({\kg_1\cdot \kg_2\over k_1 k_2}\right)^2~c(n,k_1)c(n,k_2),
\end{eqnarray}
with

\begin{eqnarray}
a(n,k)&=&{1+\sigma_8^{-0.2}(z)\left[0.7~Q_3(n)\right]^{1/2}
(q/4)^{n+3.5}\over 1+(q/4)^{n+3.5}} \cr 
b(n,k)&=&{1+0.4~(n+3)~q^{n+3}\over 1+q^{n+3.5}}\cr 
c(n,k)&=&{1+4.5/\left[1.5+(n+3)^4\right](2q)^{n+3}\over 1+(2q)^{n+3.5}},
\end{eqnarray}
and $q \equiv k/k_{NL}(z)$, where $4\pi k_{NL}^3P_L(k_{NL})=1$, and
$P_L(k)$ is the linear power spectrum at the desired redshift. The
effective spectral index is taken from the linear power spectrum as
well. The function $Q_3(n)$ is given by 

\begin{equation}
Q_3(n)={(4-2^n)\over (1+2^{n+1})},
\label{Q3def}
\end{equation}
which is the ``saturation value'' obtained in HEPT (\cite{SF99}). From
these expressions, it follows that at large scales, where the
functions $a=b=c=1$, we recover the tree-level PT. On the other hand,
at small scales, where $a^2= (7/10)Q_3(n) \sigma_8^{-0.4}$ and $b=c=0$
the bispectrum becomes hierarchical with an amplitude that
approximately reproduces HEPT for $\sigma_8 \approx 1$. For more
details see Scoccimarro \& Couchman (2000).

\section{Gravitational lensing statistics}

Calculations of gravitational lensing statistics using perturbation
theory have been presented in detail in the literature (e.g. see
\cite{BWM97}).  Therefore, we give without derivation the expressions
of the second and third moments of the convergence field $\kappa$ as
well as the skewness correction terms due to lens-lens coupling and
deviations from the Born approximation.

\subsection{Basics of Gravitational Lensing}

We use notations similar to \cite{S98} with $c=H_0=1$, since the
results do not depend on the Hubble constant. In a FRW Universe with
a matter density $\Omega_0$ and cosmological constant $\Lambda$,
the radial distance $w(z)$ and the angular diameter distance
$f_K(z)$ are defined as

\begin{equation}
w(z)=\int_0^z {{\rm d} z\over \sqrt{(1+z')^3\Omega_0+(1+z')^2
(1-\Omega_0-\Omega_\Lambda)+\Omega_\Lambda}}
\end{equation}

\begin{equation}
f_K(w)=\Big\{\matrix{\Omega_K^{-1/2}\sin{\sqrt{\Omega_K}w} ~~{\rm
for~~~ \Omega_K>0} \cr w ~~~~~~~~~~~~~~~~~~~~~{\rm for~~~ \Omega_K=0}
\cr |\Omega_K|^{-1/2}\sinh{\sqrt{|\Omega_K|}w} ~~{\rm for~~~
\Omega_K<0.} }
\end{equation}
where $\Omega_K=\Omega_0+\Omega_\Lambda-1$ is the curvature.  In the
absence of lensing, the comoving angular distance $\xg(\thetag,w)$ at
a radial distance $w(z)$ between a light ray in the direction
$\thetag$ and a fiducial light ray is $\xg(\thetag,w)=\thetag
f_K(w)$. In the presence of density fluctuations, a light ray
experiences many deflections such that $\xg(\thetag,w)$ becomes

\begin{equation}
\xg(\thetag,w)=\thetag f_K(w)-2 \int_0^w {\rm d}w'
f_K(w-w')\nablag_\perp \Phi\left[ \xg(\thetag,w'),w'\right],
\label{propeq}
\end{equation}
where $\Phi$ is the 3-dimensional gravitational potential.  The
gradient $\nablag_\perp$ denotes the derivatives with respect to $\xg$
in the plane perpendicular to the line-of-sight at distance $w$. The
amplification matrix $\Ac(\thetag,w)$ describes the mapping between
the source plane at $w(z)$

\begin{equation}
\Ac(\thetag,w)={1\over f_K(w)}{\partial \xg\over \partial\thetag},
\label{Adef}
\end{equation}
which using the first order expansion of Eq.(\ref{propeq}) becomes
$\Ac_{ij}\simeq \delta^K_{ij}+\Psi_{ij}$ with

\begin{eqnarray}
\Psi_{ij}(\thetag,w)&=& -2\int_0^w {\rm
d}w'{f_K(w-w')~f_K(w')\over f_K(w)}\times\nonumber\\
&&\Phi_{,ij}^{(1)} \left(f_K(w')\thetag,w'\right).
\label{Aapprox}
\end{eqnarray}
The convergence $\kappa$ and the shear $\gammag$ are defined such that
the amplification matrix takes the symmetric form

\begin{equation}
\Ac(\thetag,w)=\left(\matrix{1-\kappa-\gamma_1 & -\gamma_2 \cr
-\gamma_2 & 1-\kappa+\gamma_1}\right),
\end{equation}
$\kappa$ and $\gammag$ are therefore line-of-sight integrals from the
source distance $w(z)$ to $0$.  If the sources are distributed in
redshift according to some function $p_s(w)$, the convergence $\kappa$
can still be expressed as a simple line-of-sight integral

\begin{equation}
\kappa(\thetag)={3\over 2}\Omega_0\int_0^{w_H} {\rm d}w{g(w)\over
a(w)} f_K(w)~\delta(f_K(w)\thetag,w),
\label{kappadef}
\end{equation}
where

\begin{equation}
g(w)=\int_w^{w_H} {\rm d} w'~p_s(w')~{f_K(w-w')\over f_K(w')},
\label{gdef}
\end{equation}
and $w_H$ is the radial distance to the horizon.

\subsection{Variance and Skewness of the Smoothed Convergence}

In perturbation theory, the moments are calculated from a perturbative
expansion of the density contrast. The $i^{\rm th}$ order of the
convergence field is therefore

\begin{equation}
\kappa^{(i)}(\thetag)={3\over 2}\Omega_0\int_0^{w_H} {\rm
d}w{g(w)\over a(w)} f_K(w)~\delta^{(i)}(f_K(w)\thetag,w).
\label{kappapert}
\end{equation}
The calculation of the smoothed second and third moments is then
straightforward. The leading order contributions to these moments are
$\langle \kappa^2\rangle_{\theta_0}=
\langle{\kappa^{(1)}}^2\rangle_{\theta_0}$ and $\langle
\kappa^3\rangle_{\theta_0}=3
\langle{\kappa^{(1)}}^2\kappa^{(2)}\rangle_{\theta_0}$
respectively. Let us call $I(k)$ the Fourier transform of the
smoothing window and define the angular wave vector
$\sg=f_K(w)\kg_\perp$. As calculated in \cite{BWM97} and \cite{S98}
the second moment is

\begin{eqnarray}
\langle \kappa^2\rangle_{\theta_0}&=&2\pi {9\over
4}\Omega_0^2\int_0^{w_H}{\rm d}w~{g^2(w)\over a^2(w)}\times\nonumber\\
&&\int_0^\infty s{\rm d}s~P\left({s\over f_K(w)},w\right)
\left[I(s\theta_0)\right]^2,
\label{kappa2}
\end{eqnarray}
and the third moment

\begin{eqnarray}
\langle \kappa^3\rangle_{\theta_0}&=&{1\over 2\pi} {81\over 4}\Omega_0^3
\int_0^{w_H}{\rm d}w~{g^3(w)\over a^3(w)f_K(w)}\times\nonumber\\
&&\int_0^\infty{\rm d^2}\sg_1~P\left({s_1\over f_K(w)},w\right)~I(s_1\theta_0)
\times\nonumber\\
&&\int_0^\infty{\rm d^2}\sg_2~P\left({s_2\over f_K(w)},w\right)~I(s_2\theta_0)
\times \nonumber\\
&&I(|\sg_1+\sg_2|\theta_0)~F_2(\sg_1,\sg_2).
\label{kappa3}
\end{eqnarray}
These expressions rely on the validity of Eq.(\ref{Aapprox}) which
involves two approximations: one is that we assume the light rays
travel along the unperturbed light paths (the so-called Born
approximation), and the second is that we neglect non-linear terms in
the amplification matrix (which is physically equivalent to neglecting
all the couplings between subsequent lenses). These two approximations
ignore quadratic terms of the form $\kappa\cdot\kappa$,
$\kappa\cdot\gamma$ or $\gamma\cdot\gamma$ (see \cite{BWM97}). There
is no reason a priori to neglect these terms in the skewness
calculation, because they give a contribution of the same order as
Eq.(\ref{kappa3}). We shall address this question in the next Section.

\subsection{Skewness Corrections due to Lens-Lens Couplings and Deviations 
from Born Approximation}

The effect of dropping the lens-lens coupling and the Born
approximation has been calculated in the weakly non-linear regime
(see \cite{BWM97} and \cite{S98}).  The contribution to the skewness can be
calculated by considering the perturbation expansion up to the second
order in the gravitational potential in Eq.(\ref{Adef}), which leads
to a second order correction term $\Ac^{(2)}_{ij}$ in
Eq.(\ref{Aapprox}).
After lengthy but straightforward calculations, the correction to the
third moment $\langle \kappa^3\rangle_{\theta_0}^{\rm corr}$ can be
derived:

\begin{eqnarray}
\langle \kappa^3\rangle_{\theta_0}^{\rm corr}&=& {243\over
8\pi}\Omega_o^4\int_0^{w_H}{\rm d}w{g^2(w)\over a^2(w)} \int_0^w{\rm
d}w'{g(w')\over a^2(w')}{f_K(w-w')\over f_K(w)}\times\nonumber\\
&&\int{\rm d^2} \sg_1~P\left({s_1\over f_K(w)},w
\right)~I(s_1\theta)\times\nonumber\\ &&\int{\rm d^2}
\sg_2~P\left({s_2\over f_K(w')},w'
\right)~I(s_2\theta)\times\nonumber\\
&&I(|\sg_1+\sg_2|\theta)~\left[\left({\sg_1\cdot\sg_2\over s_1 s_2}
\right)^2+{\sg_1\cdot \sg_2\over s_2}\right].
\label{kappa3corr}
\end{eqnarray}
The similarity with Eq.(\ref{kappa3}) is quite obvious, which shows
that the two terms ($\langle \kappa^3\rangle_{\theta_0}$ and $\langle
\kappa^3\rangle_{\theta_0}^{\rm corr}$) are of the same order
${\delta^{(1)}}^4$ in the perturbative sense. The only difference
between the two expressions lies in the double time integral here. It
was shown (\cite{BWM97} and \cite{S98}) that this double integral
makes the correction term Eq.(\ref{kappa3corr}) only a few percent of
Eq.(\ref{kappa3}). In other words, even though the two expressions
have indeed the same order $\propto {\delta^{(1)}}^4$, they differ by
almost one order of magnitude just because of the additional
$f_K(w-w')/f_K(w)$ factor in Eq.(\ref{kappa3corr}) (in the case of a
single source redshift this factor is equivalent to an additional
$g(w)$ factor, according to Eq.(\ref{gdef})). It is therefore correct
to say that the lensing efficiency function $g(w)$ can play the role
of a perturbative parameter as the density contrast does (when $\delta
\ll 1$).  The smallness of $\langle \kappa^3\rangle_{\theta_0}^{\rm
corr}$ shows that even in the weakly non-linear regime, terms
involving powers of $g(w)$ higher than $1$ will always be small
compared to terms only proportional to $g(w)$. In the following we
shall compute Eq.(\ref{kappa3corr}) in the non-linear regime, and use
results from ray-tracing simulations in order to check if this
statement is still valid when the density contrast is much larger than
$1$.

\subsection{Calculations and Filtering Schemes}

We can now calculate the variance, the skewness and its correction of
the convergence field using the Eq.(\ref{kappa2}), (\ref{kappa3}) and
(\ref{kappa3corr}) by forming the ratios:

\begin{equation}
S_3(\kappa)={\langle \kappa^3\rangle_{\theta_0}\over
\langle \kappa^2\rangle_{\theta_0}^2} \ \ \ ; \ \ \
\Delta S_3(\kappa)={\langle \kappa^3\rangle_{\theta_0}^{\rm corr}\over
\langle \kappa^2\rangle_{\theta_0}^2}
\label{skewnessdef}
\end{equation}
It is worth to notice that $\Delta S_3(\kappa)$ is weakly dependent on
the cosmological parameters since the $\Omega_0$ dependence in the
front of Eq.(\ref{kappa2}) and Eq.(\ref{kappa3corr}) cancels out. The
non-linear variance is given by Eq.(\ref{kappa2}) where the power
spectrum is replaced by the Peacock \& Dodds (1996) prescription as in
Jain \& Seljak (1997). The non-linear third moment is given by Eq.(\ref{kappa3})
where $F_2(\kg_1,\kg_2)$ has been replaced by $F^{\rm
eff}_2(\kg_1,\kg_2)$ and we use the non-linear power spectrum.  For
the skewness correction, we only need to use the non-linear power
spectrum in Eq.(\ref{kappa3corr}).

We use two types of filtering schemes: top-hat filtering and $M_{\rm
ap}$ statistics. The Fourier transform of the top-hat window is:

\begin{equation}
I(\eta)={{\rm J}_1(\eta)\over \pi\eta},
\end{equation}
where $J$ denotes a Bessel function. The $M_{\rm ap}$ statistic can
be defined either from the convergence $\kappa$ or from the tangential
shear $\gamma_t$:

\begin{equation}
M_{\rm ap}=\int {\rm d}^2\thetag~U(\theta) \kappa(\thetag)=
\int {\rm d}^2\thetag~Q(\theta) \gamma_t(\thetag),
\label{mapdef}
\end{equation}
where $\theta^2 Q(\theta)=
2~\int_0^\theta~{\rm d}\theta'~\theta'~U(\theta')-\theta^2 U(\theta)$,
and $U(\theta)$ is the compensated filter. We use the compact compensated filter
described in Schneider et al. (1998) where $U(\theta)$ is defined only for
$\theta\in [0,~\theta_0]$, and

\begin{equation}
U(\theta)={9\over \pi \theta_0^2} \left[1-\left({\theta\over \theta_0}\right)^2\right]
\left[{1\over 3}-\left({\theta\over \theta_0}\right)^2\right].
\end{equation}
The Fourier transform of $U(\theta)$ is

\begin{equation}
I(\eta)={12~{\rm J}_4(\eta)\over \pi\eta^2},
\end{equation}
which is the function to be inserted in Eqs.(\ref{kappa2},
\ref{kappa3}, \ref{kappa3corr}).
The motivation for using the compensated filters is
illustrated on Fig.2 of Schneider et al. (1998): this filter is a pass-band
filter, which is a more direct estimate of the mass band powers than the
top-hat filter.
Another advantage of the compensated filter is that the measurement of the
skewness does not require a mass reconstruction (whereas top-hat filtered
statistic does) as it can be evaluated directly from the shear
(Eq.(\ref{mapdef})). Note also that the compensated filter defined above
peaks at a scale $\sim \theta_0/5$, if $\theta_0$ is the radius of the
filter. Therefore at equal radius with the top-hat filter, $M_{\rm ap}$
statistics probes scales as small as $\sim \theta_0/5$.

\section{Ray-tracing simulations}

We use two different kind of ray-tracing simulations (high and low
resolution simulations) which allow us to investigate scales from
$0.1'$ to several tens of arcmins. The high resolution simulations
($0.1'$) were done by Jain, Seljak \& White (2000), and are
described in that paper. The investigated models in those simulations
are OCDM2 and $\tau$CDM (see Table \ref{table}). The area covered by
each simulation is $9$ square degrees.

The numerical methods used for the low resolution simulation are
basically same as those of the high resolution simulation except for a
use of a tiled set of the independent particle-mesh (PM) $N$-body
simulations\footnote{This code, developed by the {\em Numerical
Investigations in Cosmology} group ({\em N.I.C.}), is an improved
version of an earlier work by Bouchet, Adam \& Pellat (1985) and by
Moutarde et al.~(1991), fully vectorized and optimized to run in
parallel on several processors of a NEC-SX5.}.  The PM simulations use
$256^2\times512$ particles and a force mesh of the same size in a
periodic rectangular comoving box, with outputs done along the
light-cone (Hamana, Colombi and Suto 2000).  We performed 10, 11 and
12 independent simulations to generate the density field from $z\sim
2$ to the present for SCDM, OCDM1 and $\Lambda$CDM models,
respectively.  We use the {\it tiling} technique first proposed by
White \& Hu (2000), i.e., the box size of each simulation is chosen so
as to match to the convergence of the light ray bundle.  The smallest
box size is comoving $80\times80\times160h^{-3}$Mpc$^3$ for three
models, while the largest box size are comoving
$240\times240\times480h^{-3}$Mpc$^3$,
$320\times320\times640h^{-3}$Mpc$^3$ and
$400\times400\times800h^{-3}$Mpc$^3$, for SCDM, OCDM1 and $\Lambda$CDM
models, respectively.  These tiled sets of simulations yield a
field-of-view of $5\times5$ square degrees.

The multiple lens-plane algorithm is used for ray-tracing as in Jain
et al.~(2000).  The lens planes are located between $z=0$ and $z\sim
2$ at intervals of comoving $80h^{-1}$Mpc.  We obtained 40
realizations by randomly shifting the simulation boxes.  For each
realization, $512^2$ rays are traced backward from the observer's
point.  The initial ray directions are set on $512^2$ grids with a
grid spacing of $5^\circ/512\sim 0.59$ arcmin.  Although the results
of the ray-tracing simulations (the position and amplification matrix
of each rays) are stored at all lens planes, in this paper we only
focus on the results at the plane closest to $z=1$. The error bars on
the results of the low resolution simulations correspond to the
standard deviation among 40 realizations.
\begin{table}
\caption{Ray-tracing simulations used in this work. The SCDM, OCDM1
and $\Lambda$CDM are described in Hamana et al. (in preparation) and
the OCDM2 and $\tau$CDM are from Jain, Seljak \& White
(2000). $\theta_{res}$ indicates the angular resolution of the
simulations $N_{rea}$ indicates the number of realizations per model,
and Area specifies the sky coverage for each realization in square
degrees. Note that the $N_{rea}=40$ realizations are not rigorously
independent, being random rotations of a single suite of simulations.}
\begin{center}
\begin{tabular}{|c|c|c|c|c|c|}\hline
Simulations & SCDM & OCDM1 & $\Lambda$CDM & OCDM2 & $\tau$CDM \\ \hline\hline
$\Omega_0$ & 1 & 0.3 & 0.3 & 0.3 & 1 \\
$\Lambda$ & 0 & 0 & 0.7 & 0 & 0 \\
$\Gamma$ & 0.5 & 0.21 & 0.21 & 0.21 & 0.21 \\
$\sigma_8$ & 0.6 & 0.85 & 0.9 & 0.85 & 0.6 \\
$z_s$ & 1.0025 & 1.03 & 1.0034 & 1 & 1 \\
$\theta_{res}$ & 2' & 2' & 2' & 0.1' & 0.1' \\
$N_{rea}$ & 40 & 40 & 40 & 7 & 5 \\
$ {\rm Area} $ & 25 & 25 & 25 & 9 & 9 \\
\hline
\end{tabular}
\end{center}
\label{table}
\end{table}

The range of filtering scales where the results of the low resolution
simulation are reliable is between $4'$ and $30'$ for the statistics
of the top-hat filtered convergence with source redshift $z\sim1$.  On
the other hand, for the $M_{ap}$ statistics, that is between $20'$ and
$100'$: the reason is that our compensated filter probes scales as
small as $\sim \theta_0/5$ (see Section 3.4), therefore $M_{ap}$
predictions are likely to become inaccurate below $4'\times 5=20'$.
The small-scale resolution limit comes from the finite spatial
resolution of our PM simulations, while the large-scale limit comes
from a lack of power in the density field on scales larger than the
simulation box (finite size effects).

\begin{figure*}
\epsfysize=3.7in
\centerline{\epsffile{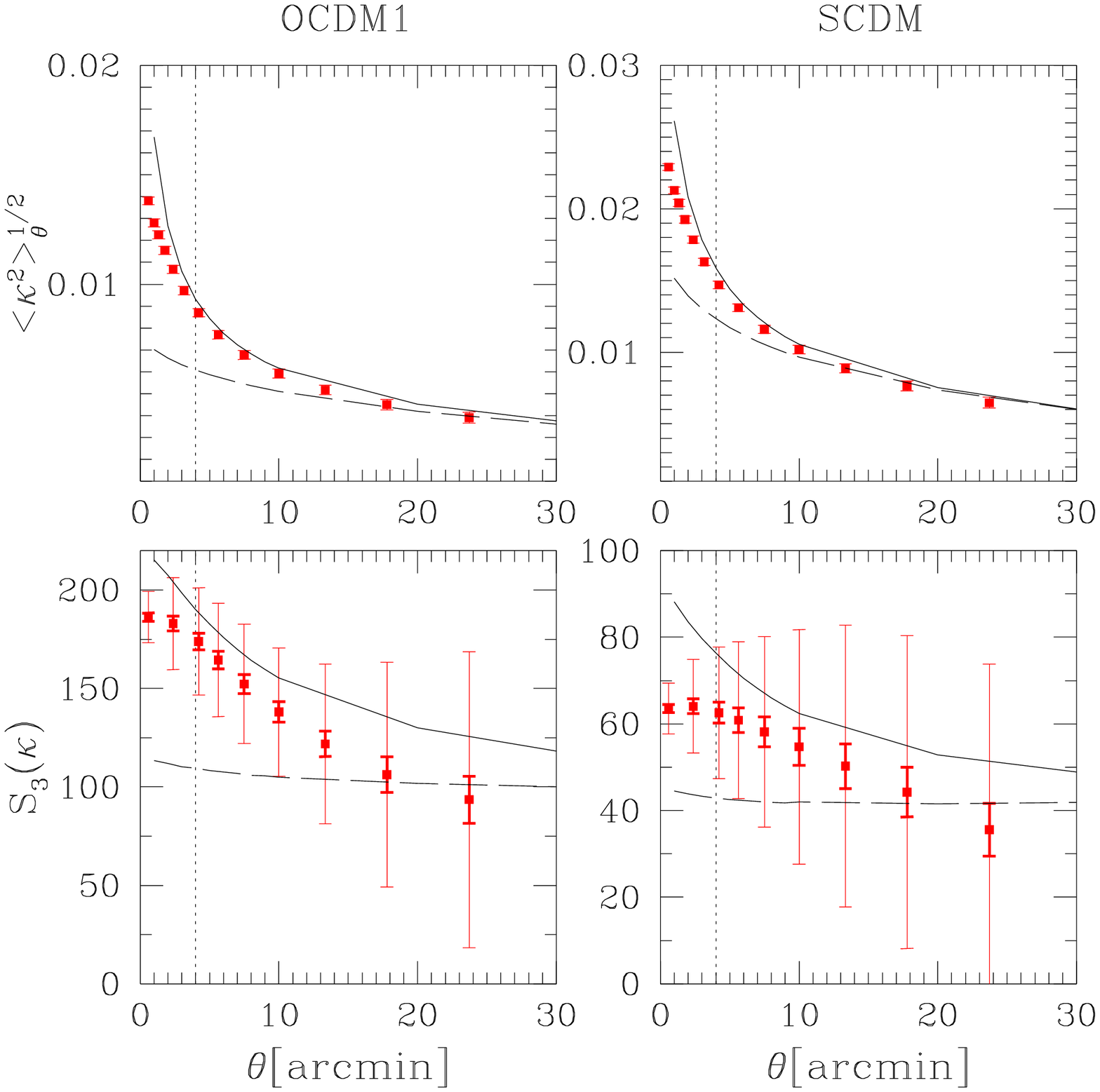}\epsfysize=3.7in\epsffile{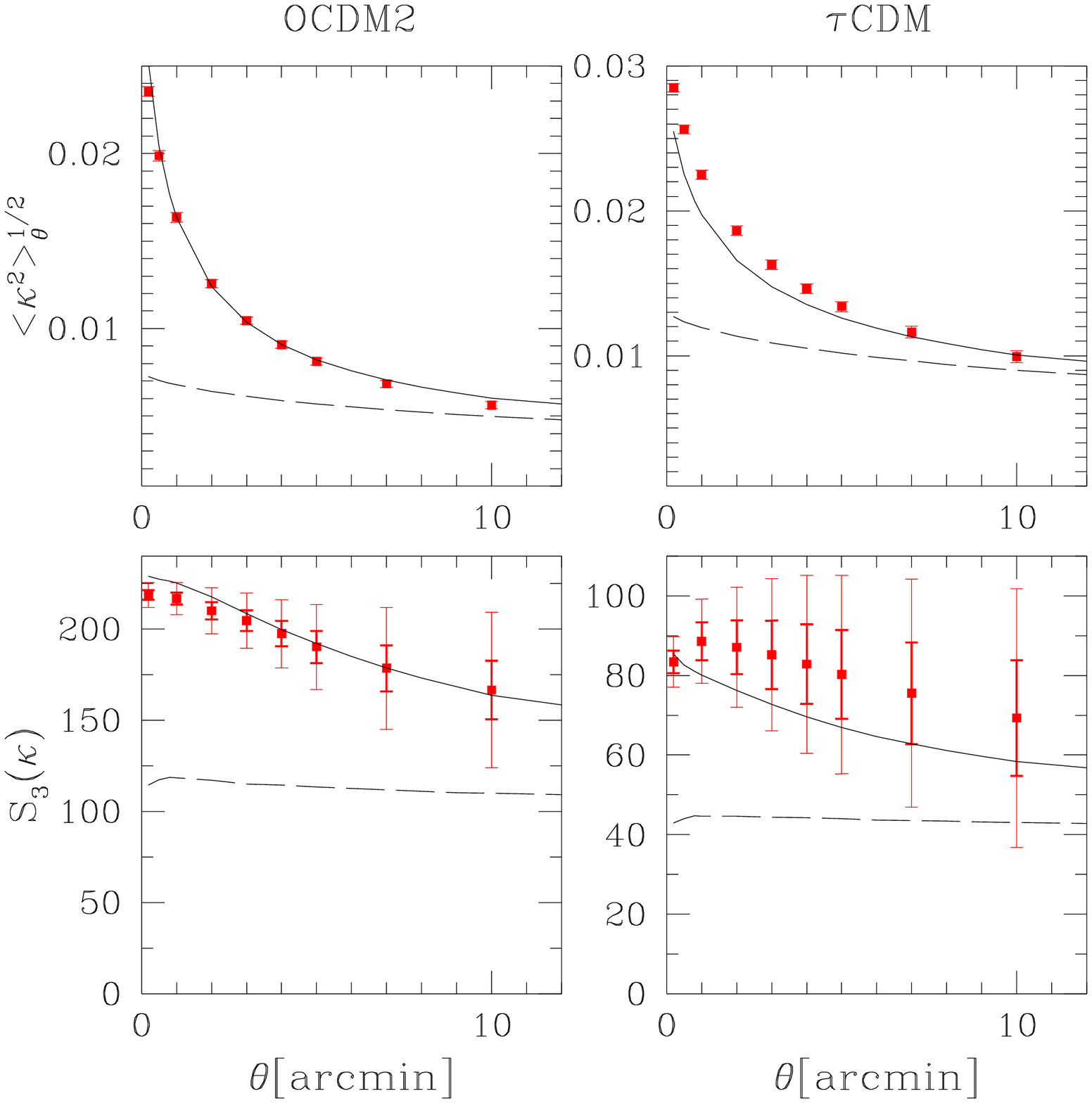}}
\caption{\label{TH.ps} r.m.s. (top panels) and skewness (bottom
panels) of the convergence field as a function of top-hat smoothing
scale for the OCDM1 (far left panels) and SCDM (center-left panels)
OCDM2 (center-right panels) and $\tau$CDM (far right panels) models
given in Table \ref{table}.  The solid line is the prediction using
non-linear prescriptions for the power-spectrum and the bispectrum,
and the dashed lines show the perturbation theory calculations. The
vertical dotted lines in the left panels denote the reliable scale
limit. The ``large'' error bars correspond to scaling the ``small''
error bars (thick lines) to a single 25 square degree (left panels)
and 9 square degree (right panels) survey.}
\end{figure*}

In addition to the usual {\it full} ray-tracing simulations, we also
performed {\it approximated} ray-tracing which neglects both the
deflection of the ray trajectory and the couplings between subsequent
lenses.  To be more specific, the shear tensor, $\bm{U}_m$ (see eq.~13
of Jain et al.~2000 for a definition), is computed on the {\it
unperturbed} ray position, and the distortion tensor $\bm{\Phi}_m$ in
the right-hand-side of the eq.~(14) of Jain et al. (2000) is replaced
with the identity matrix to neglect all lens-lens couplings.  The weak
lensing statistics are then computed in the same manner as in the {\it
full} ray-tracing case.  The difference in the skewness obtained from
the {\it full} and {\it approximated} simulations are computed to test
the validity of the perturbative treatment in the semi-analytic
predictions (\S 5).

\section{Results}

\begin{figure*}
\epsfysize=3.7in
\centerline{\epsffile{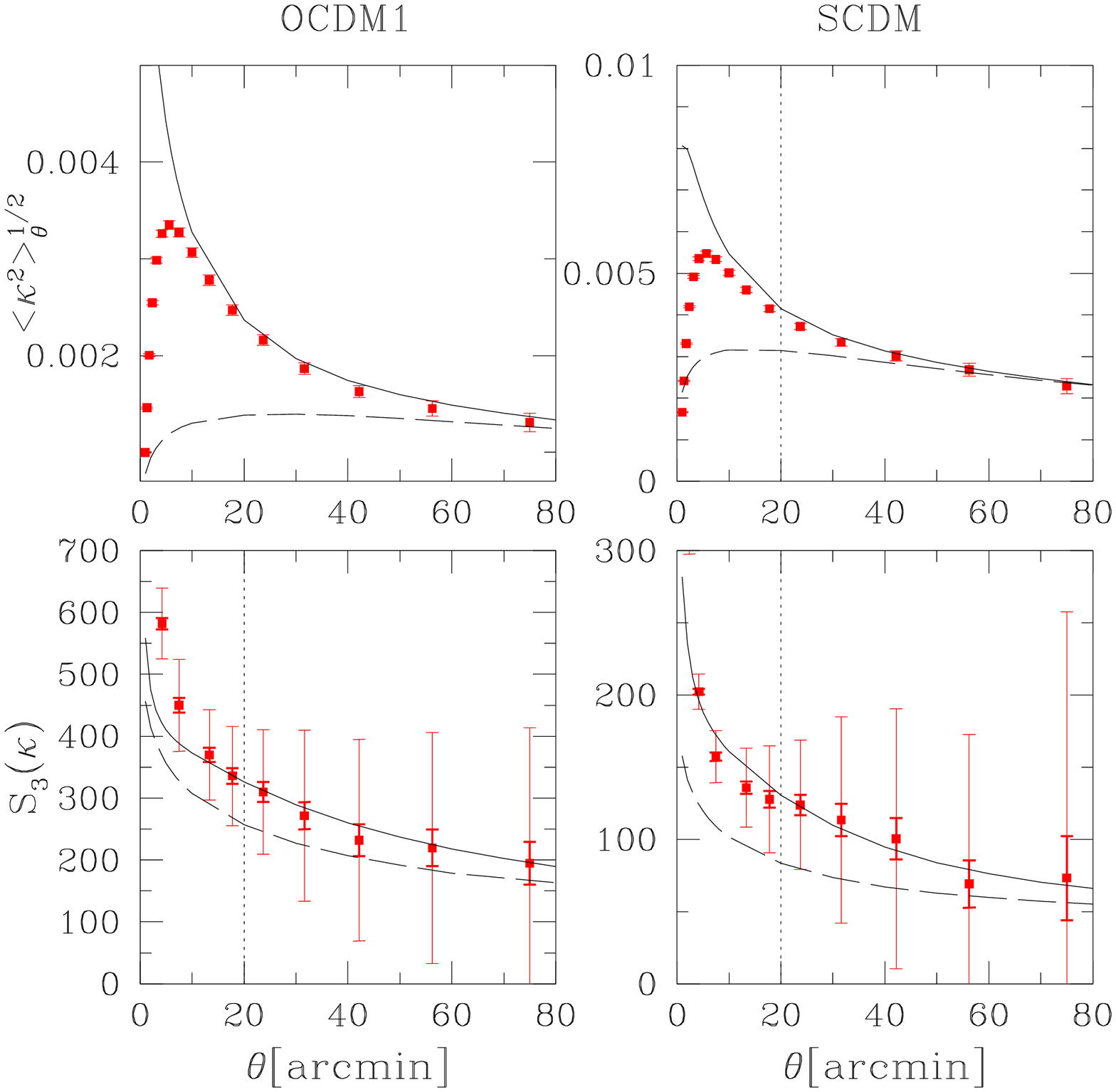}\epsfysize=3.7in\epsffile{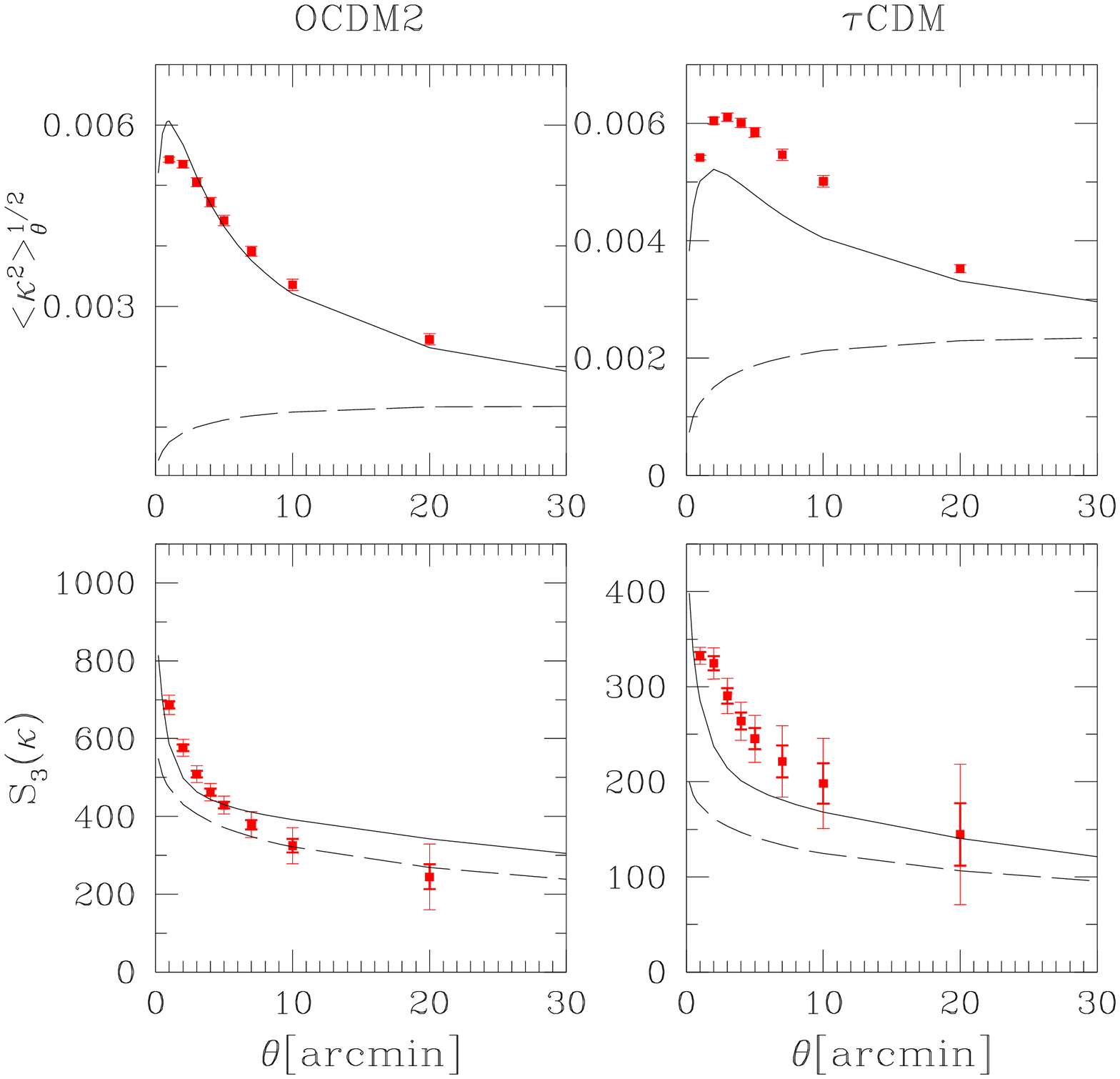}}
\caption{\label{MAP.ps} Same as Figure \ref{TH.ps}, but for the
compensated filter.}
\end{figure*}

Figures \ref{TH.ps} and \ref{MAP.ps} show the measured r.m.s. and
skewness of the convergence compared to the linear (dashed lines) and
non-linear (solid lines) predictions. The square symbols represent the
measurements done in the simulations. Two error bars are drawn per
point: the small error bars correspond to the dispersion in the
measurements taken into account all the realizations, whereas the
large error bars are obtained by scaling (by $\sqrt{N_{rea}}$) the
small error bars to a single realization, relevant for assesing cosmic
variance in a single field of the corresponding area.

For the top-hat filtering scheme (Figure \ref{TH.ps}) the non-linear
predictions give very good results for all the scales investigated
(i.e. down to scales as small as $0.1'$).  For the low resolution
simulation (Figure \ref{TH.ps} left panels), agreement starts to
deteriorate once we reach the reliable scale limit which is $4'$
(shown by the vertical dotted lines). At smaller scales the high
resolution simulations results show that the fitting formula still
works down to $0.1-0.2'$.
The agreement starts to deteriorate quickly above the scale of $\sim 30'$
because finite field effects become important, therefore these scales
are not shown here. Moreover, from a practical point
of view we are less interested in these scales because residual systematics
from shear measurements are supposed to dominate the lensing
signal (Erben et al. 2000). The few percent discrepancy for the
r.m.s. for the $\tau$CDM model (top right of Figure \ref{TH.ps} right
panels) was already observed by Jain, Seljak \& White (2000), but this
is consistent with the accuracy of the power spectrum fitting formula
(\cite{PD96}). The non-linear predictions for the skewness in
Fig.~\ref{TH.ps}, although consistent within the errors with
ray-tracing simulations, shows fluctuations of $S_3(\kappa)$ by
approximately $10\%$. This is within the expected accuracy of the
bispectrum fitting formula ($15\%$, \cite{SC00}); however, one must
also keep in mind that this formula was obtained from single
realizations of CDM models, and the bispectrum amplitude is rather
sensitive to the presence of massive clusters (see Cooray \& Hu 2000
for the effects of this on $S_3(\kappa)$), therefore the residual
$10\%$ fluctuation might be due to cosmic variance.

For the compensated filter (Figure \ref{MAP.ps}), the difference
between linear and non-linear variance is more significant, since when
the spectral index becomes $n<-2$ the variance in a compensated field
starts to decrease (\cite{S98}), and this happens at much larger
scales for the linear than the non-linear power spectrum. The
agreement between predictions and simulations is very good for the low
resolution simulations, including the skewness, and good for OCDM2,
the high resolution case. On the other hand, the agreement is not as
good for the high resolution $\tau$CDM model. The likely explanation
is that the small discrepancy observed in that case for the top hat
filter (Fig. \ref{TH.ps}) is amplified by the compensated filter,
which acts like a derivative operator on the convergence field: as a
pass-band filter, the compensated filter is more sensitive to any
small discrepancy (in $k$-bands) between the non-linear power-spectrum
fitting formula and the ray-tracing simulations, whereas the top-hat
filter integrates over large range of $k$'s and tends to mask such
discrepancies. We should note that the effects of resolution also
cause the variance to decrease at small scales (top left panels in
Fig.~\ref{MAP.ps}), therefore, at scales smaller than the resolution
the agreement between the skewness prediction and the measurement in
the low resolution simulation (bottom left panels in
Fig.~\ref{MAP.ps}) is coincidental.

\begin{figure*}
\epsfysize=3.7in
\centerline{\epsffile{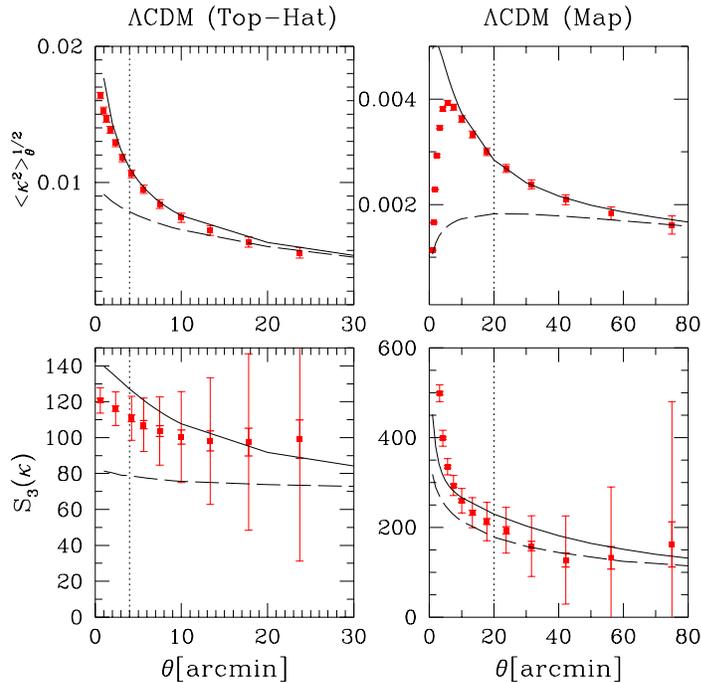}}
\caption{\label{LCDM.ps} Top-hat (left panels) and compensated (right panels)
statistics for the $\Lambda$CDM model (table \ref{table}).}
\end{figure*}

\begin{figure}
\epsfysize=3.in
\epsffile{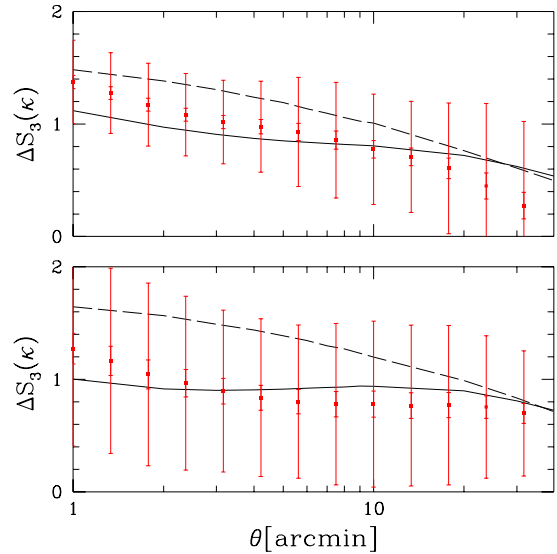}
\caption{\label{delta_s3.ps} Skewness correction for top-hat smoothing
$\Delta S_3(\kappa)$ as defined in Eq.(\ref{skewnessdef}) for the
model SCDM (top panel) and OCDM1 (bottom panel). Squares with error
bars correspond to measurements in the ray-tracing simulations, dashed
lines to the prediction of perturbation theory, and solid lines to
predictions from the non-linear fitting formula.}
\end{figure}

Figure \ref{delta_s3.ps} shows the skewness correction for top-hat
smoothing in SCDM (top panel) and OCDM1 (bottom panel) models. The
dashed line and the solid line show respectively the predictions from
perturbation theory and the non-linear fitting formula, whereas points
with error bars denote measurements in the ray-tracing
simulations. Surprisingly, even in the non-linear regime, the
correction remains small and, in fact, the relative skewness
correction is smaller than in the weakly non-linear regime!  As
explained in Section 3.3, this means that coupling terms introduce a
non-linearity which is always smaller than the non-linearity produced
by very dense (non-linear) structures. In other words, even in the
non-linear regime, the dominant lensing contribution comes from the
single lens-plane approximation, which physically means that the
probability for a given line-of-sight to pass through two very dense
structures at very different resdhifts is very low. Similar results
hold for the $M_{\rm ap}$ statistics (not shown here) where, in fact,
the correction terms to the skewness are even smaller.

\begin{figure*}
\epsfysize=3.in
\centerline{\epsffile{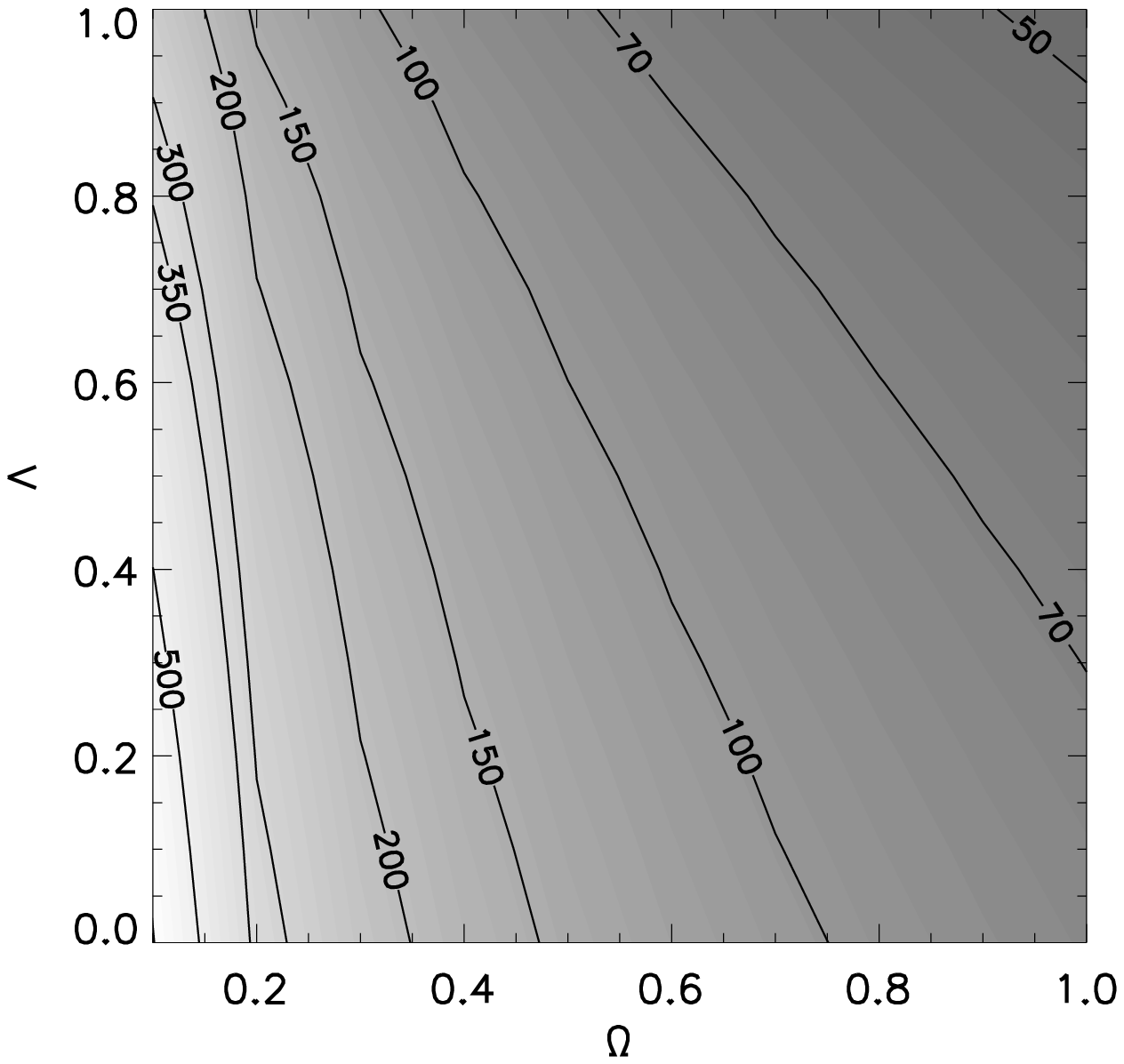}\epsfysize=3.in\epsffile{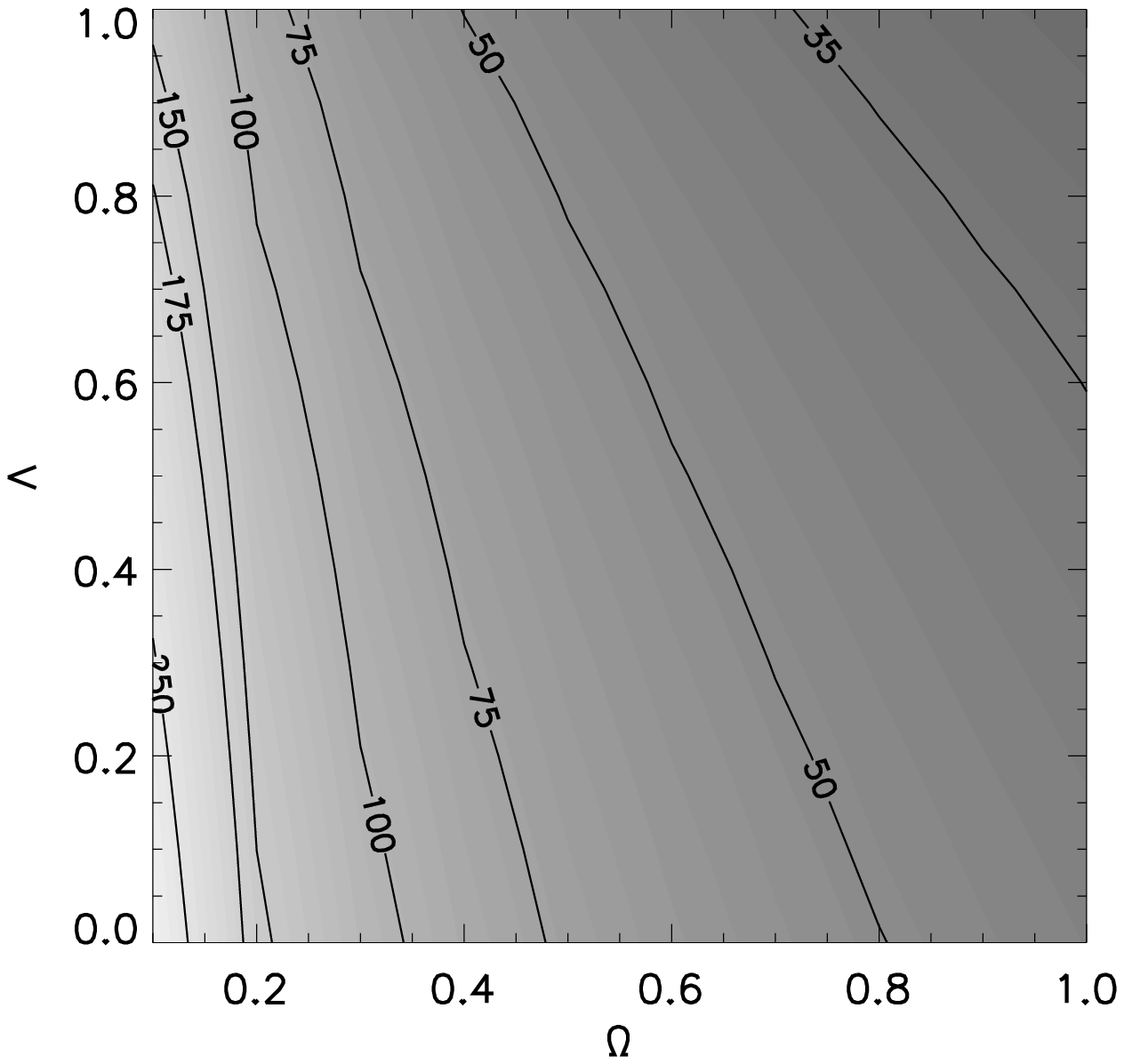}}
\caption{\label{skew_om_lam.ps} The expected skewness $S_3(\kappa)$ in
the non-linear regime at a scale of $1'$ (left) compared to
$S_3(\kappa)$ computed using perturbation theory at a scale of $10'$
(right) in the $\Omega$-$\Lambda$ plane. Top-Hat filtering was
used. The sources are located at redshift unity. Note that for a given
$\Omega$-$\Lambda$ the skewness in the left panel is roughly twice the
value on the right panel.}
\end{figure*}

Figure \ref{skew_om_lam.ps} shows the predicted skewness $S_3(\kappa)$
in the $\Omega-\Lambda$ plane. Here we compare small scale (1')
non-linear calculations (left panel) to larger scale (10')
perturbation theory calculations (right panel) for top-hat
smoothing. Note that the choice of a scale for the perturbative 
estimate is almost not relevant since it does depend very weakly on
scale (Figure \ref{TH.ps}).  This Figure shows that the cosmological
dependence is preserved in the non-linear regime, and that
$S_3(\kappa)$ remains essentially an estimator of $\Omega$ (the
degeneracy along $\Lambda$ is strong). This result has to be compared
with the Cosmic Microwave Background and the supernovae constraints,
which are sensitive to the curvature and $\sim\Lambda$
respectively. This clearly motivates further lensing studies as an
independent means of measuring the cosmological parameters. In fact
the lensing constraints are similar to constraints derived from the
galaxy catalogues or the velocity fields, although the physics is much
simpler here, and weak lensing does not suffer from the galaxy bias
problem.

The skewness predictions in the non-linear regime are twice their
value using perturbation theory at large scales. Therefore the cosmic
variance and noise analysis done in Van Waerbeke, Bernardeau \&
Mellier (1999) underestimated the capability of lensing surveys by a
significant amount, essentially because they used very low resolution
and approximate dynamics in their simulations\footnote{In this work
the authors used very simple lensing simulations where the mass
distribution was modeled with second-order 2-dimensional lagrangian
dynamics, which they had to smooth at small scale in order to avoid
singularities.  Because of this smoothing they obtained a skewness of
the convergence close to its value in perturbation theory.}. In this
work the authors shown that a $25$ square degrees survey would measure
the skewness at $9\sigma$, in standard observational conditions, if
$\Omega=0.3$ ($3\sigma$ if $\Omega=1$).  Here, the signal amplitude is
twice larger, therefore we would need $\sim 3-4$ times less sky
coverage to reach the same signal-to-noise ratio to measure
$S_3(\kappa)$. However, the exact gain depends on the precise amount
of cosmic variance and shot noise at this scale.  Since the cosmic
variance is already given in Figure~\ref{TH.ps}, we can simply add the
shot noise obtained in Van Waerbeke, Bernardeau \& Mellier (1999) to
the high resolution simulation results presented here.  It turns out
that the shot noise is roughly $\sigma_{S_3}\sim 10$ at $1-2$ arcmin,
therefore we can reach the signal-to-noise ($7-9\sigma$ for OCDM) with
only $9$ square degrees with a single measurement at $1'$.  However we
should keep in mind that our results for the error on the skewness in
the $\Lambda$CDM model are consistent although slightly smaller than
previous estimates from White \& Hu (2000), possibly due to the fact
that our realizations are not completely independent (being random
samples of a single suite of simulations).

\begin{figure}
\epsfysize=3.in
\epsffile{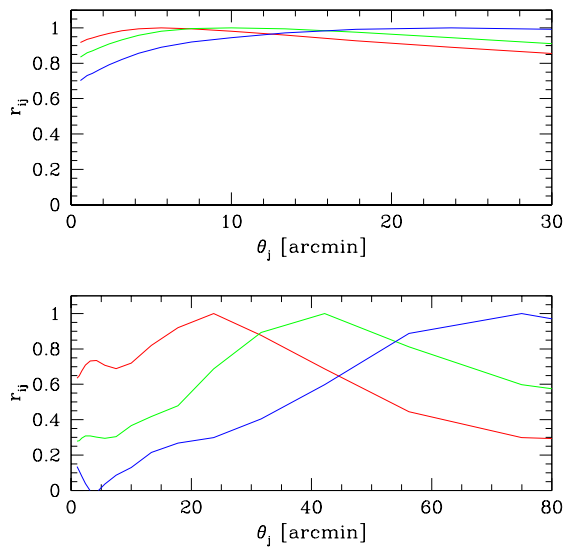}
\caption{\label{rij_var.eps} The cross-correlation coefficient
$r_{ij}$ between the variance measured at angular scale $\theta_i$ and
$\theta_j$, for the $\Lambda$CDM model. The top panel shows $r_{ij}$
for top-hat filtering at $\theta_i=5.6,10,23.7$ arcmin as a function
of $\theta_j$. The bottom panel shows $r_{ij}$ for compensated filters
at $\theta_i=23.7,42.2,75$ arcmin as a function of $\theta_j$.}
\end{figure}

Finally, Fig.~\ref{rij_var.eps} shows the cross-correlation
coefficient $r_{ij}$ between the variance measured at angular scale
$\theta_i$ and $\theta_j$, for the $\Lambda$CDM model (similar results
hold for the other cosmological models and for the skewness
measurements as well). These results were obtained by averaging over
the 40 realizations in our ensemble. The top panel presents results
for the top hat filter, whereas the bottom panel corresponds to the
compensated filter. We show three angular scales,
$\theta_i=5.6,10,23.7$ arcmin for the top hat case, and
$\theta_i=23.7,42.2,75$ arcmin for the compensated filter. The top hat
measurements of the variance show a large cross-correlation, in fact,
the cross-correlation coefficient is almost unity among all scales. On
the other hand, as expected from being a band-pass filter, the
compensated filter measurements are much less correlated. Note however
than this result only includes cosmic variance contributions, i.e. it
does not include shot noise, which can be a significant source of
cross-correlations at small scales.

\section{Conclusions}

We calculated semi-analytically the r.m.s. and the skewness of the
convergence field $\kappa$ and showed that the non-linear prescription
for the 3-dimensional bispectrum given in Scoccimarro \& Couchman
(2000) describes the transition from perturbative to non-linear
angular scales very well with an accuracy $\sim 10\%$.  We combined
different ray-tracing simulations with largely different resolutions
($0.1'$ and $2'$) to compare our predictions. The transition from the
weakly non-linear scales to the non-linear regime is clearly visible
and also accurately described by our semi-analytical approach for both
top-hat and compensated smoothing schemes for the OCDM, SCDM and
$\Lambda$CDM models (The $\tau$CDM predictions are not very accurate
for the compensated filter, but we suggest that it might come from
innacuracies in the non-linear power spectrum fitting formula).  

Our results extend previous non-linear calculations of the skewness of
$\kappa$ which either assumed the validity of HEPT at all redshifts
(\cite{H99}) or relied on the hierarchical clustering ansatz
(\cite{V00,MC00,BV00}) at intermediate scales. Our semi-analytical
approach allows us to investigate the cosmological dependence of the
skewness of $\kappa$ in as many models as desired, without having to
perform a ray tracing simulation each time.

By comparing the skewness in ray-tracing simulations with and without
lens-lens coupling and Born approximation terms, we demonstrated that
the single lens-plane approximation remains valid even in the highly
non-linear regime, which is a non-trivial statement about weak lensing
at small scales. We found that the contribution of these terms to the
skewness is in remarkable agreement with the semi-analytical
predictions.

We have also calculated $S_3(\kappa)$ in the $\Omega-\Lambda$ plane
and shown that the signal is about twice the expected value from
perturbation theory with roughly the same degeneracy in $\Lambda$ as
in the perturbative case, generalizing previous results (Jain et
al. 2000, White \& Hu 2000). In addition, our results on the cosmic
variance of the skewness suggests that a combined analysis of the
existing weak lensing surveys (\cite{VW00,WT00,BRE00,KWL00}) should
already be able to measure the skewness of the convergence.

\section{acknowledgments}
FB and LvW thank IAS for hospitality, where part of this work was
done.  We thank Bhuvnesh Jain for the invaluable use of his
ray-tracing simulations and for useful comments on the manuscript.  RS
is supported by endowment funds from the Institute for Advanced Study
and NSF grant PHY-0070928 at IAS.  This research was supported in part
by the Direction de la Recherche du Minist{\`e}re Fran{\c c}ais de la
Recherche.  The computational means (CRAY-98 and NEC-SX5) to do the
$N$-body simulations were made available to us thanks to the
scientific council of the Institut du D\'eveloppement et des
Ressources en Informatique Scientifique (IDRIS). Numerical computation
in this work was partly carried out at the TERAPIX data center and on
MAGIQUE (SGI-O2K) at IAP.

\end{document}